\documentclass[twocolumn,english]{article}
\usepackage[T1]{fontenc}
\usepackage{babel}
\usepackage{units}
\usepackage{amsmath}
\usepackage{amssymb}
\usepackage{graphicx}
\usepackage[unicode=true,
 bookmarks=true,bookmarksnumbered=true,bookmarksopen=true,bookmarksopenlevel=1,
 breaklinks=false,pdfborder={0 0 0},backref=false,colorlinks=false]
 {hyperref}
\hypersetup{pdftitle={Your Title},
 pdfauthor={Your Name},
 pdfpagelayout=OneColumn, pdfnewwindow=true, pdfstartview=XYZ, plainpages=false}

\makeatletter

\providecommand{\tabularnewline}{\\}

\pdfoutput=1

\makeatother

\begin{document}

\title{Numerical Stability Improvements for the Pseudo-Spectral EM PIC Algorithm}

\author{Brendan B. Godfrey,%
\thanks{Brendan Godfrey is with the University of Maryland, College Park,
Maryland 20742, USA, e-mail: \protect\href{mailto:brendan.godfrey@ieee.org}{brendan.godfrey@ieee.org}.%
} Jean-Luc Vay,%
\thanks{Jean-Luc Vay is with Lawrence Berkeley National Laboratory, Berkeley,
California 94720, USA, e-mail: \protect\href{http://jlvay@lbl.gov}{jlvay@lbl.gov}.%
} and~Irving Haber%
\thanks{Irving Haber is with the University of Maryland, College Park, Maryland
20742, USA, e-mail: \protect\href{mailto:haber@umd.edu}{haber@umd.edu}.%
}}
\maketitle
\begin{abstract}
The pseudo-spectral analytical time-domain (PSATD) particle-in-cell
(PIC) algorithm solves the vacuum Maxwell's equations exactly, has
no Courant time-step limit (as conventionally defined), and offers
substantial flexibility in plasma and particle beam simulations. It
is, however, not free of the usual numerical instabilities, including
the numerical Cherenkov instability, when applied to relativistic
beam simulations. This paper presents several approaches that, when
combined with digital filtering, almost completely eliminate the numerical
Cherenkov instability. The paper also investigates the numerical stability
of the PSATD algorithm at low beam energies.
\end{abstract}

\section{Introduction}

The Pseudo Spectral Analytical Time Domain (PSATD) particle-in-cell
(PIC) algorithm solves Maxwell's equations in spatial Fourier space,
using equations that are exact for any time step, when the plasma
current is constant \cite{HaberICNSP73}. Consequently, it has no
Courant limit in the usual sense and, when combined with the Esirkepov
conserved current algorithm \cite{esirkepov2001exact}, is highly
accurate for relativistic beam simulations. The PSATD algorithm nonetheless
has received little attention over the past two decades, perhaps due
to the belief that it could not be used efficiently on modern parallel
computers. Recently, however, an approach to PSATD parallelization
has been developed, and simulations performed to demonstrate its effectiveness
\cite{Vay2013PSATD}.

The numerical stability properties of the PSATD-Esirkepov algorithm
have been analyzed to develop methods of suppressing the all too pervasive
numerical Cherenkov instability \cite{godfrey1974numerical}. Results
included a complete multidimensional numerical dispersion relation
for cold relativistic beams, approximate analytical expressions for
numerical instability peak growth rates, and numerical solutions of
the dispersion relation for various choices of numerical parameters
\cite{Godfrey2013PSATD}. Importantly, this analysis demonstrated
that numerical Cherenkov instability growth rates could be reduced
by as much as two orders of magnitude by suitable choices of algorithm
options and digital filtering.

In this paper still other approaches for suppressing the numerical
Cherenkov instability are introduced. These new options require less
digital filtering than before, preserving a larger portion of k-space
for physical phenomena. Additionally, analysis of the numerical Cherenkov
instability is extended to lower energies, where it eventually gives
way to the well known electrostatic numerical instability \cite{Langdon1970,Okuda1970}
for non-relativistic beams. An instability unique to pseudo-spectral
algorithms, first predicted in \cite{BirdsallLangdon}, also is discussed.

\section{Prior Stability Results}

Although multiple representations of the PSATD algorithm exist, the
following is both compact and convenient for numerical analysis:
\begin{multline}
\mathbf{E}^{n+1}=\mathbf{E}^{n}-2iS_{h}\mathbf{k}\times\mathbf{B}^{n+\nicefrac{1}{2}}/k-S\mathbf{\mathbf{\zeta}:}\mathbf{J}_{e}^{n+\nicefrac{1}{2}}/k\\
+S\mathbf{k}\mathbf{k}\cdot\mathbf{\mathbf{\zeta}:}\mathbf{J}_{e}^{n+\nicefrac{1}{2}}/k^{3}-\mathbf{k}\mathbf{k}\cdot\mathbf{J}_{e}^{n+\nicefrac{1}{2}}\triangle t/k^{2},\label{eq:leapfrog alt1}
\end{multline}
\begin{equation}
\mathbf{B}^{n+\nicefrac{3}{2}}=\mathbf{B}^{n+\nicefrac{1}{2}}+2iS_{h}\mathbf{k}\times\mathbf{E}^{n+1}/k,\label{eq:leapfrogB}
\end{equation}
with $\mathbf{B}^{n}$, needed to advance the particles, given by
\begin{equation}
\mathbf{B}^{n}=\frac{1}{2C_{h}}\left(\mathbf{B}^{n+\nicefrac{1}{2}}+\mathbf{B}^{n-\nicefrac{1}{2}}\right).\label{eq:Bave}
\end{equation}
Quantities introduced in these and subsequent equations are given
in \cite{Godfrey2013PSATD}, upon which this section of the paper
is based. 

Lengthy algebra, validated with Mathematica \cite{Mathematica9},
leads to a dispersion relation of the form,
\begin{multline}
C_{0}+n\sum_{m_{z}}C_{1}\csc\left[\left(\omega-k_{z}^{\prime}v\right)\frac{\Delta t}{2}\right]+\\
n\sum_{m_{z}}\left(C_{2x}+\gamma^{-2}C_{2z}\right)\csc^{2}\left[\left(\omega-k_{z}^{\prime}v\right)\frac{\Delta t}{2}\right]\\
+\gamma^{-2}n^{2}\left(\sum_{m_{z}}C_{3z}\csc^{2}\left[\left(\omega-k_{z}^{\prime}v\right)\frac{\Delta t}{2}\right]\right)\\
\left(\sum_{m_{z}}C_{3x}\csc\left[\left(\omega-k_{z}^{\prime}v\right)\frac{\Delta t}{2}\right]\right)=0,\label{eq:drformfull}
\end{multline}
with $k_{z}^{\prime}=k_{z}+m_{z}\,2\pi/\Delta z$ and $m_{z}$ an
integer. The factors, $\csc\left[\left(\omega-k_{z}^{\prime}v\right)\frac{\Delta t}{2}\right]$,
except that multiplying $C_{3z}$, are numerical artifacts. Coupling
of the resulting numerical modes, $\omega=k_{z}^{\prime}v$, with
electromagnetic modes, $C_{0}=0$, causes the numerical Cherenkov
instability.

\begin{figure}[tbh]
\begin{centering}
\includegraphics[scale=0.6]{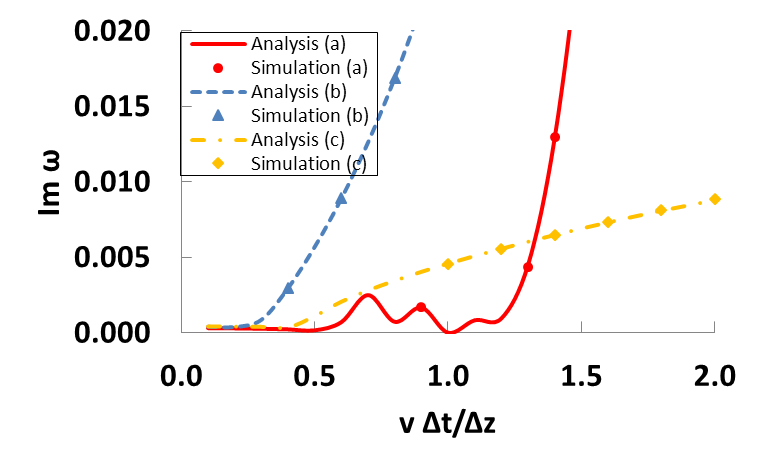}
\par\end{centering}

\caption{\label{fig:cubic-filtered}Maximum growth rates for options (a), (b),
and (c) of Table \ref{tab:Tableoption} with cubic interpolation and
the digital filter in Eq. \ref{eq:filter10}. Markers represent corresponding
simulation results.}

\end{figure}

The strategy used in \cite{Godfrey2013PSATD} to suppress the relativistic
beam simulation numerical Cherenkov instabilities is as follows. The
fastest growing instabilities occur where the $m_{z}=0,\,-1$ alias
modes intersect the light cone, typically at large $k$, where they
can be eliminated by digital filtering. The higher order aliases often
are unstable at small \emph{k}, but have small growth rates that can
be reduced even further by using higher order interpolation, say,
cubic. Finally, the $m_{z}=0,\,-1$ aliases often are unstable, although
with smaller growth rates, well away from the light cone, typically
as small \emph{k}. Higher order interpolation has only modest impact
on the non-resonant instability associated with the $m_{z}=-1$ alias,
and almost no impact on that associated with the $m_{z}=0$ alias.
Fortunately, they can be reduced in growth rate by appropriate choices
for $\zeta$, such as those listed in Table \ref{tab:Tableoption},
with $\zeta_{x}=\zeta_{z}=$ 
\begin{equation}
\frac{k\, k_{z}\frac{\triangle t}{2}\left(\sin^{2}\left(k_{z}\frac{\Delta t}{2}\right)-\sin^{2}\left(k\frac{\Delta t}{2}\right)\right)\csc\left(k_{z}\frac{\Delta t}{2}\right)\csc\left(k\frac{\Delta t}{2}\right)}{k_{z}\sin\left(k_{z}\frac{\Delta t}{2}\right)\cos\left(k\frac{\Delta t}{2}\right)-k\,\cos\left(k_{z}\frac{\Delta t}{2}\right)\sin\left(k\frac{\Delta t}{2}\right)}\label{eq:resz}
\end{equation}
when $k_{z}<\frac{\pi}{\Delta t}-k_{x}^{2}\frac{\Delta t}{4\pi}$
and 0 otherwise for option (c). Peak growth rates obtained using this
strategy with cubic interpolation and the ten-pass digital filter
given by 
\begin{multline}
\cos^{16}\left(k_{z}\frac{\Delta z}{2}\right)\left(5-4\cos^{2}\left(k_{z}\frac{\Delta z}{2}\right)\right)^{2}\\
\cos^{16}\left(k_{x}\frac{\Delta x}{2}\right)\left(5-4\cos^{2}\left(k_{x}\frac{\Delta x}{2}\right)\right)^{2}\label{eq:filter10}
\end{multline}
and depicted in Fig. 7 of \cite{Godfrey2013PSATD} are shown in Fig.
\ref{fig:cubic-filtered}. Growth rates for options (a) or (c) are
sufficiently small that the numerical Cherenkov instability can safely
be ignored in most applications when $v\,\Delta t/\Delta z<1.3$.
Growth rate curves were obtained by solving Eq.\ref{eq:drformfull}
for the $m_{z}=+1,\,0,\,-1$ aliases, and the simulation results using
the WARP-PSATD PIC code\cite{Warp}. In Figs. \ref{fig:cubic-filtered}-\ref{fig:nulodd},
$\gamma=130$.

\begin{table}[tbh]
\begin{centering}
\begin{tabular}{|c|c|}
\hline 
Option & Current Factors for Eq. \ref{eq:leapfrog alt1}\tabularnewline
\hline 
\hline 
(a) & %
\begin{tabular}{c}
$\zeta_{x}=\left(k_{x}\triangle x/2\right)\cot\left(k_{x}\triangle x/2\right)$\tabularnewline
$\zeta_{z}=\left(k_{z}\triangle z/2\right)\cot\left(k_{z}\triangle z/2\right)$\tabularnewline
\end{tabular}\tabularnewline
\hline 
(b) & $\zeta_{z}=\zeta_{x}=1$\tabularnewline
\hline 
(c) & $\zeta_{x}=\zeta_{z}$, as defined in Eq. \ref{eq:resz}\tabularnewline
\hline 
\end{tabular}
\par\end{centering}

\caption{\label{tab:Tableoption}Algorithm options used in Fig. \ref{fig:cubic-filtered}
and elsewhere.}
\end{table}

\section{High Gamma Results}

\begin{figure}[tbh]
\begin{centering}
\includegraphics[scale=0.6]{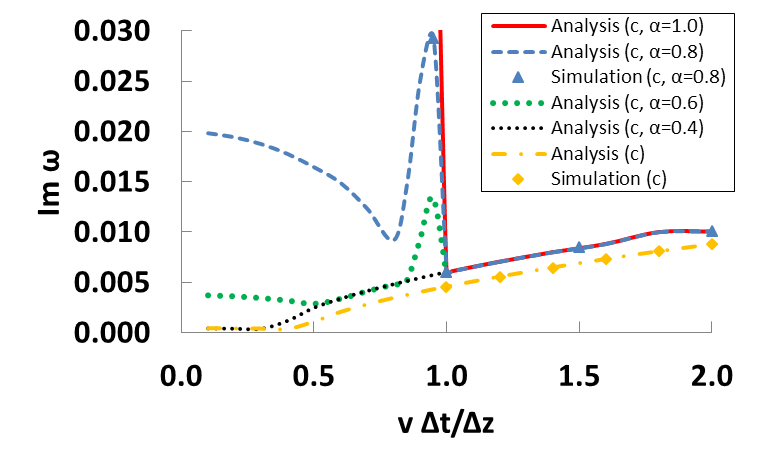}
\par\end{centering}

\caption{\label{fig:Sharp cutoff}Maximum growth rates for option (c) of Table
\ref{tab:Tableoption} with cubic interpolation and sharp-cutoff digital
filter with threshold values $\alpha=0.4,\,0.6,\,0.8,\,1.0$. Markers
represent corresponding simulation results.}
\end{figure}

The digital filter carried over from \cite{Godfrey2013PSATD} and
employed in obtaining Fig. \ref{fig:cubic-filtered} is quite aggressive,
discarding about 70\% of k-space in each dimension. (It was chosen
to facilitate comparison with an earlier Finite Difference Time Domain
(FDTD) numerical Cherenkov instability analysis \cite{godfrey2013esirkepov}
and with FDTD production simulations of Laser Plasma Acceleration
\cite{VayJCP2011,VayPoP2011}.) The effect of less aggressive filtering,
in this case a sharp cutoff at $\alpha\,\min\left[\frac{\pi}{v\,\triangle t},\frac{\pi}{\triangle z}\right]$
is illustrated in Fig. \eqref{fig:Sharp cutoff} for $\alpha=1.0,\,0.8,\,0.6,\,0.4$.
The option (c) results from Fig. \eqref{fig:cubic-filtered} are reproduced
for comparison. Growth rates are scarcely affected by the choice of
filtering for $v\triangle t/\triangle z>1,$ because option (c) is
so effective at suppressing the $m_{z}=0$ instability, which otherwise
would dominate there. However, option (c) has only modest effect on
the $m_{z}=-1$ instability, which typically dominates for $v\triangle t/\triangle z<1$.
As is evident from the figure, the choice of $\alpha$ matters greatly
there, and $\alpha\lesssim0.7$ is necessary to keep the numerical
Cherenkov instability in check . Nonetheless, the sharp-cutoff digital
filter at least doubles the proportion of usable k-space for $v\triangle t/\triangle z\approx1$,
compared to the filter in Eq. \eqref{eq:filter10}.

\begin{figure}[tbh]
\begin{centering}
\includegraphics[scale=0.6]{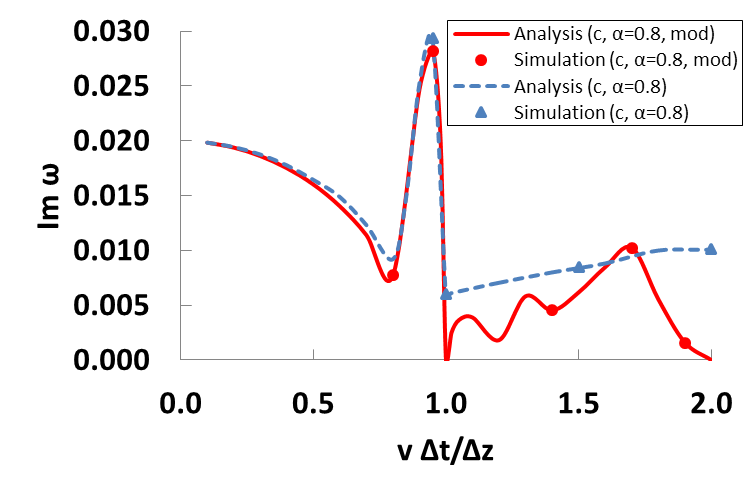}\caption{\label{fig:Optc-mod}Maximum growth rates for option (c) of Table
\ref{tab:Tableoption} with cubic interpolation and $\alpha=0.8$
sharp-cutoff digital filter. The ``mod'' curve is obtained by modifying
$\zeta_{z}$ as described in the text. Markers represent corresponding
simulation results.}

\par\end{centering}

\end{figure}

Incidentally, the maximum instability growth at larger $v\triangle t/\triangle z$
in Figs. \eqref{fig:cubic-filtered} and \eqref{fig:Sharp cutoff}
scales roughly as $\gamma^{-1}$ but can be eliminated in part by
making $\zeta_{z}$ for option (c) look more like $\zeta_{z}$ for
option (a) at small \emph{k}, for instance, by multiplying it by $\sin^{\nicefrac{1}{6}}\left(k_{z}\triangle z/2\right)$.
(This odd expression was obtained by numerical trial and error.) Fig.
\ref{fig:Optc-mod} shows the effect of this factor for option (c)
with $\alpha=0.8$. The residual growth for $v\triangle t/\triangle z>1$
is from higher order aliases and could be reduced further by using
higher order interpolation, if desired.

\begin{figure}[tbh]
\begin{centering}
\includegraphics[scale=0.6]{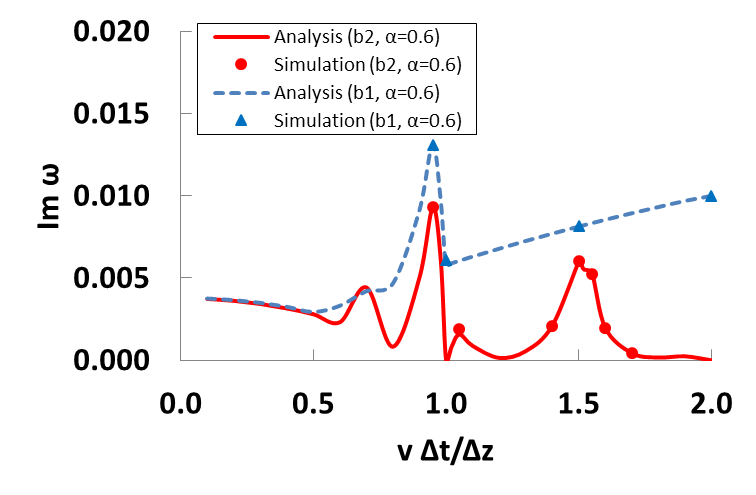}\caption{\label{fig:eb12}Maximum growth rates for field multiplier options
(b1) and (b2) described in the text, cubic interpolation, $\zeta_{z}=\zeta_{x}=1$,
and $\alpha=0.6$ sharp-cutoff digital filter. Markers represent corresponding
simulation results.}

\par\end{centering}

\end{figure}

\begin{figure}[tbh]
\begin{centering}
\includegraphics[scale=0.6]{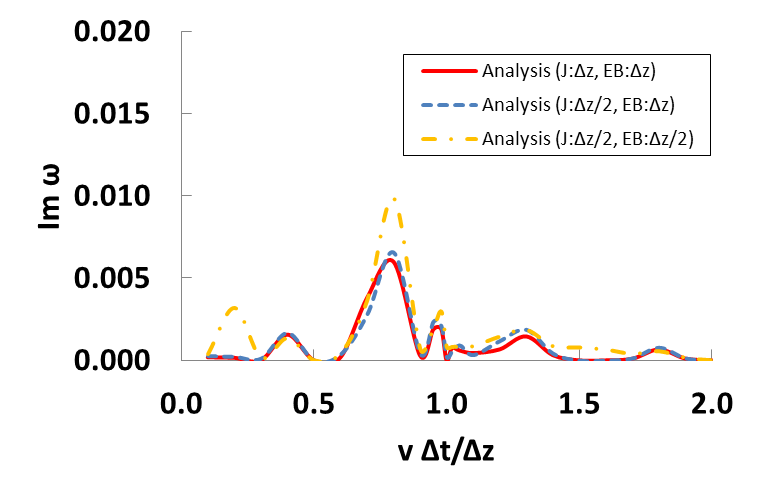}
\par\end{centering}

\caption{\label{fig:nulodd}Maximum growth rates for odd-$m_{z}$ elimination
procedures described in the text, cubic interpolation, option (c)
modified, and $\alpha=0.8$ sharp-cutoff digital filter. }

\end{figure}

The numerical Cherenkov instability can be viewed as arising from
slight mismatches among the electric and magnetic fields due to finite
differencing. Because peak growth rates tend to scale as the one-third
power of the transverse force, even quite small errors are sufficient
to produce the instabilities depicted in the preceding figures. Introducing
the current scaling terms, $\mathbf{\zeta}$, in Eq. \ref{eq:leapfrog alt1}
helps to reduce these errors. An alternative, more direct approach
is to rescale slightly the transverse fields as seen by the particles.
For instance, choosing a multiplier for $E_{x}$,
\begin{equation}
\frac{2\frac{k}{k_{z}v}\sin^{2}\left(\frac{k\,\triangle t}{2}\right)\sin\left(\frac{k_{z}v\triangle t}{2}\right)\cos\left(\frac{k_{z}v\triangle t}{2}\right)}{k\,\triangle t\sin^{2}\left(\frac{k\,\triangle t}{2}\right)+\left(\sin\left(k\,\triangle t\right)-k\,\triangle t\right)\sin^{2}\left(\frac{k_{z}v\triangle t}{2}\right)}\label{eq:eb1}
\end{equation}
causes the term $C_{2x}$ in Eq. \ref{eq:drformfull} to vanish to
order $\gamma^{-2}$ at $\omega=k_{z}v$, thereby reducing the order
of the highest order pole in the dispersion relation from 2 to 1 in
the high-$\gamma$ limit. The Fig. \ref{fig:eb12} curve labeled ``b1''
gives results for this multiplier, cubic interpolation, option (b)
current factor, and $\alpha=0.6$ sharp-cutoff digital filter. It
is not surprising that the option (b1) curve strongly resembles the
option (c), $\alpha=0.6$ curve in Fig. \ref{fig:Sharp cutoff}, because
option (c) also is designed to zero $C_{2x}$ at $\omega=k_{z}v$
\cite{Godfrey2013PSATD}.

The second curve in Fig. \ref{fig:eb12}, labeled ``b2'', is based
on an alternative pair of multipliers for $E_{x}$ and $B_{y},$ respectively
\begin{equation}
\frac{k_{z}v\triangle t}{2}\cot\left(\frac{k_{z}v\triangle t}{2}\right),\;\frac{k\triangle t}{2}\cot\left(\frac{k\triangle t}{2}\right),
\end{equation}
chosen to causes the term $C_{3x}$ in Eq. \ref{eq:drformfull} to
vanish at $\omega=k_{z}v$, reducing the order of the highest order
pole in the dispersion relation from 3 to 2. It is very effective
at suppressing the numerical Cherenkov instability.

Note that an $\alpha=0.6$ cutoff is used for options (b1) and (b2)
to avoid a weak residual numerical Cherenkov instability that occurs
at larger $k_{z}$ and small $k_{x}$. This instability occasionally
has been observed in the PSTD algorithm, described in Sec. 7 of \cite{Godfrey2013PSATD},
and in unpublished work by the first author on a version of PSATD
employing vector potentials. It can be described approximately by
\begin{equation}
C_{0}+n\sum_{m_{z}}C_{2x}\csc^{2}\left[\left(\omega-k_{z}^{\prime}v\right)\frac{\Delta t}{2}\right]=0\label{eq:Residual}
\end{equation}
for large $\gamma$. It appears to be triggered by the close proximity
in \emph{k}-space of the $m=0$ numerical beam mode and one of the
electromagnetic modes over a range of large $k_{z}$ values when $k_{x}$
is small. (No $m=0$ instability occurs at $k_{x}=0$, however.) Attempts
apart from digital filtering to suppress this remaining $m=0$ instability
have been unsuccessful.

All the preceding methods for suppressing the numerical Cherenkov
instability rely on digital filtering to curtail the $m_{z}=-1$ alias.
At some cost in additional computation, it is straightforward to eliminate
odd-numbered aliases entirely. For instance, one can simply double
the number of cells in \emph{z} while cutting $\Delta z$ in half,
then discard the upper half of $k_{z}$-space (suggested by Birdsall
and Maron for electrostatic simulations \cite{Birdsall1980}). This,
coupled with cubic interpolation and option (c) modified, yields the
maximum growth rate curve labeled ``$\mathrm{\mathsf{J:\triangle z/2,\, EB:\triangle z/2}}$'',
which is excellent. Of course, this approach approximately doubles
the run time of a simulation. More economical is to compute currents
as just described but compute fields and interpolate them to the particles
on the original, coarser mesh. This second approach increases the
run time of the original simulation by only about 50\% and produces
modestly better stability properties besides. See the curve labeled
``$\mathrm{\mathsf{J:\triangle z/2,\, EB:\triangle z}}$''. Going
one step further, one could compute the currents on the refined mesh
as before but using splines based on a $\triangle z$ rather than
a $\triangle z/2$ unit cell. However, doing so achieves negligible
improvement in run time or stability, and the curve labeled ``$\mathrm{\mathsf{J:\triangle z,\, EB:\triangle z}}$''
is included only for completeness. For all three curves in this figure,
Eq. \ref{eq:drformfull} was solved for the $m_{z}=+2,\,0,\,-2$ aliases.

\section{Low Gamma Results}

\begin{figure}[tbh]
\begin{centering}
\includegraphics[scale=0.6]{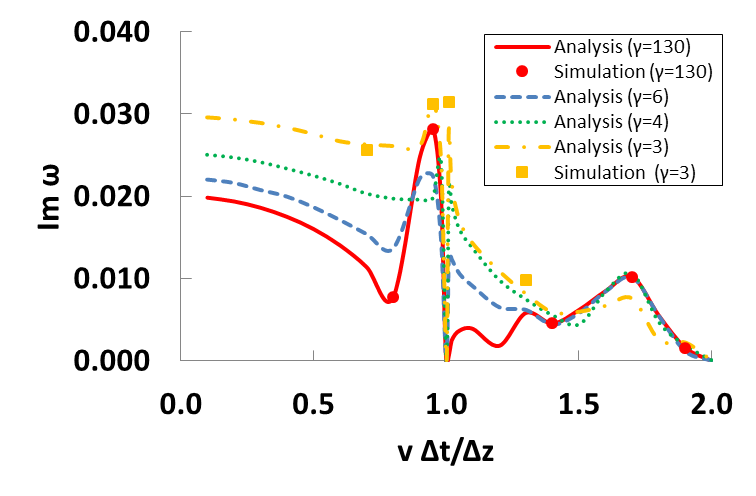}
\par\end{centering}

\caption{\label{fig:vargam}Maximum growth rates for $\gamma=130,\,6,\,4,\,3$,
with cubic interpolation, option (c) modified, and $\alpha=0.8$ sharp-cutoff
digital filter. Markers represent corresponding simulation results.}

\end{figure}

Although the focus of this and the two preceding papers on suppressing
the numerical Cherenkov instability \cite{Godfrey2013PSATD,godfrey2013esirkepov}
is on highly relativistic beams, appropriate to Laser Plasma Accelerator
simulations \cite{VayPoP2011,VayPoPL2011}, examining the scaling
of numerical instabilities properties with $\gamma$ is of interest
both to validate the numerical dispersion relation of Eq. \ref{eq:drformfull}
over a range of beam energies and also to determine the robustness
of proposed methods to suppress numerical instabilities. Fig 10 of
\cite{Godfrey2013PSATD} effectively substantiated the dispersion
relation itself, successfully comparing predicted instability growth
rates with those measured in WARP for $\gamma=130,\,3.0,\,1.4,\,1.1$.
Fig. \ref{fig:vargam} illustrates how peak growth rates vary with
$\gamma$ for one of the best instability suppression methods, option
(c) modified with cubic interpolation and the $\alpha=0.8$ sharp-cutoff
digital filter. Instability growth rates, essentially independent
of beam energy for $\gamma\gtrsim10$, steadily increase for smaller
beam energies due to the emergence of a well known, predominantly
$m_{z}=-1$, non-relativistic, electrostatic numerical instability
\cite{Langdon1970,Okuda1970}. The instability suppression method
employed in Fig. \ref{fig:vargam} is, evidently, progressively less
effective at lower beam energies, although it still is able to reduce
peak growth by an order of magnitude for $\gamma$ as small as 2.

\begin{figure}[tbh]
\begin{centering}
\includegraphics[scale=0.6]{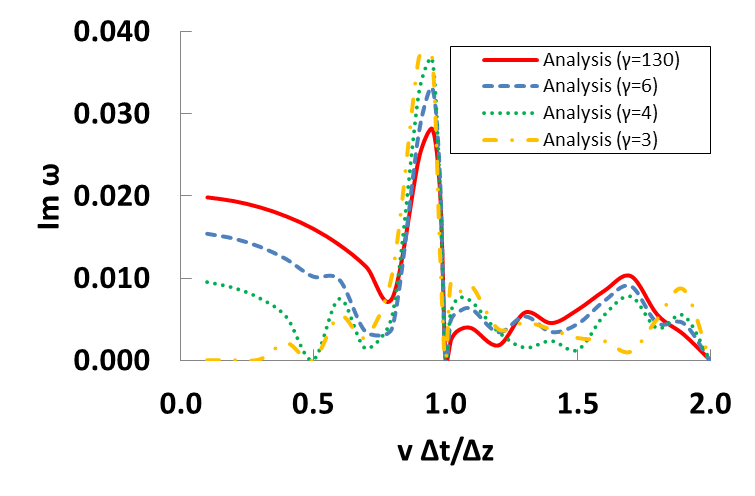}
\par\end{centering}

\caption{\label{fig:vargamoffset}Maximum growth rates for $\gamma=130,\,6,\,4,\,3$,
with $E_{z}$ offset, cubic interpolation, option (c) modified, and
$\alpha=0.8$ sharp-cutoff digital filter. }

\end{figure}

Birdsall and Langdon years ago recommended ``energy-conserving''
algorithms \cite{BirdsallLangdon,Birdsall1980,lewis1972variational,Langdon1973energy}
for controlling the electrostatic numerical instability. Many variants
are available. For the PSATD-Esirkepov algorithm the simple approach
of offsetting $E_{z}$, the electric field aligned with the beam,
by one-half cell in \emph{z} works well, as illustrated in Fig. \ref{fig:vargamoffset}.
The ``Galerkin algorithm'' described in \cite{godfrey2013esirkepov},
another energy-conserving algorithm, works not quite as well, not
because it is ineffective against the electrostatic numerical instability
but because it is less effective against the numerical Cherenkov instability
\cite{Godfrey2013PSATD}. Not surprisingly, eliminating odd aliases,
as described in the last section, also is quite effective at suppressing
the numerical electrostatic instability.

\begin{figure}[tbh]
\begin{centering}
\includegraphics[scale=0.6]{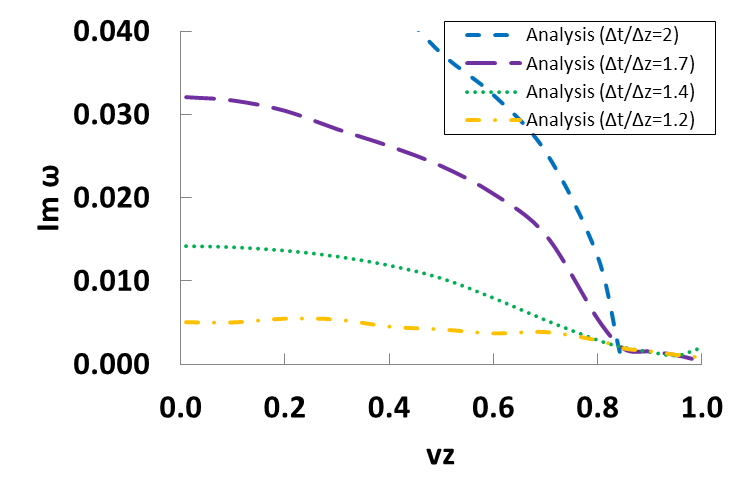}
\par\end{centering}

\caption{\label{fig:lowgam}Maximum growth rates for $\triangle t/\triangle z=2.0,\,1.7,\,1.4,\,1.2$
with cubic interpolation, $\mathrm{\mathsf{J:\triangle z,\, EB:\triangle z}}$
odd alias elimination, option (c) modified, and no digital filter.}

\end{figure}

The numerical instability suppression techniques described in this
article are less effective against another instability, predicted
in Problem 15-9a of \cite{BirdsallLangdon}, which occurs for $m_{z}=0$
at very low $\gamma$. It satisfies the dispersion relation,
\begin{multline}
\sin^{2}\left(\omega\frac{\Delta t}{2}\right)-\sin^{2}\left(k\frac{\Delta t}{2}\right)\\
-n\left(S^{J}\right)^{2}\frac{\zeta_{z}\Delta t}{4k}\sin\left(k\Delta t\right)=0,\label{eq:BL}
\end{multline}
obtained in a straightforward manner from Eq. \ref{eq:drformfull}
with $v=0$ and $\zeta_{z}=\zeta_{x}$. (This dispersion relation
is related to Eq. (59) in \cite{Godfrey2013PSATD}, which is valid
for all $\gamma$ but only at $k_{x}=0$.) Eq. \ref{eq:BL} has complex
roots in typically very narrow bands in k-space located at just less
than $k=l\,\pi/\triangle t$, \emph{l} an integer. Fig. \ref{fig:lowgam}
shows that the maximum growth rate of this instability gradually decreases
as the beam velocity increases. Growth of this instability vanishes
for $v\gtrsim0.8$, although growth rates from weaker instabilities
remain visible in the figure. These calculations were performed for
$m_{z}=+2,\,0,\,-2$ aliases with cubic interpolation, $\mathrm{\mathsf{J:\triangle z,\, EB:\triangle z}}$
odd alias elimination, option (c) modified, and no digital filter.
Without such aggressive suppression of the numerical Cherenkov and
electrostatic instabilities, the instability in question would not
have been visible. Even so, all three instabilities have comparable
maximum growth rates for $\triangle t/\triangle z=1.2$.

\section{Conclusions}

This paper introduces a variety of methods for suppressing the numerical
Cherenkov instability in highly relativistic beam PSATD-Esirkepov
PIC simulations complementing the methods introduced in \cite{Godfrey2013PSATD}.
Option (a) is both simple and highly effective for suppressing the
numerical Cherenkov instability when $v\,\Delta t/\Delta z<1.3$,
at least for the parameters used in this paper. For larger values
of $v\,\Delta t/\Delta z$, option (c) modified with $\alpha=0.8$
seems best on balance. Option (b2) has modestly better stability properties
but requires $\alpha=0.6$ to avoid a new, not yet fully explained
instability. The odd-alias elimination techniques also are highly
effective, although at the cost of increased run time.

Numerical instability growth rates are nearly independent of beam
energy for $\gamma\gtrsim10$. At lower energies, the well known numerical
electrostatic instability gradually increases in importance, becoming
dominant for $\gamma\lesssim3$. Offsetting $E_{z}$ by one-half cell
in \emph{z} is reasonably effective at controlling the electrostatic
instability. The odd-alias elimination techniques seem highly effective
here as well. Also to be considered at low energies is the numerical
instability predicted by Birdsall and Langdon \cite{BirdsallLangdon}
for pseudo-spectral algorithms, which occurs at \emph{k} an integer
multiple of $\pi/\triangle t$. Although fairly weak and highly localized
in \emph{k}-space, this instability apparently cannot be suppressed
except by strongly digitally filtering the tiny region of \emph{k}-space
in which it occurs. 

A new instability unpredictably occurs in some instances in a small
region of \emph{k}-space located at large $k_{z}$ and small $k_{x}$.
Obtaining a better understanding of it is a priority of the first
author.

Most of the results presented in this paper were obtained using Mathematica
\cite{Mathematica9}. The authors intend to publish the Mathematica
program used to produce the results in this article on a publicly
accessible web site, http://hifweb.lbl.gov/public/BLAST/Godfrey/.

\section*{Acknowledgment}

We thank David Grote for assistance with the code WARP. This work
was supported in part by the Director, Office of Science, Office of
High Energy Physics, U.S. Dept. of Energy under Contract No. DE-AC02-05CH11231
and the US-DOE SciDAC ComPASS collaboration, and used resources of
the National Energy Research Scientific Computing Center.

\bibliographystyle{IEEEtran}
\bibliography{IEEEabrv,IEEEexample,C:/Users/Brendan/Documents/LyX/Biblio_JCP_Godfrey,C:/Users/Brendan/Documents/LyX/Godfrey}

\begin{thebibliography}{10}
\providecommand{\url}[1]{#1}
\csname url@samestyle\endcsname
\providecommand{\newblock}{\relax}
\providecommand{\bibinfo}[2]{#2}
\providecommand{\BIBentrySTDinterwordspacing}{\spaceskip=0pt\relax}
\providecommand{\BIBentryALTinterwordstretchfactor}{4}
\providecommand{\BIBentryALTinterwordspacing}{\spaceskip=\fontdimen2\font plus
\BIBentryALTinterwordstretchfactor\fontdimen3\font minus
  \fontdimen4\font\relax}
\providecommand{\BIBforeignlanguage}[2]{{%
\expandafter\ifx\csname l@#1\endcsname\relax
\typeout{** WARNING: IEEEtran.bst: No hyphenation pattern has been}%
\typeout{** loaded for the language `#1'. Using the pattern for}%
\typeout{** the default language instead.}%
\else
\language=\csname l@#1\endcsname
\fi
#2}}
\providecommand{\BIBdecl}{\relax}
\BIBdecl

\bibitem{HaberICNSP73}
I.~Haber, R.~Lee, H.~Klein, and J.~Boris, ``Advances in electromagnetic
  simulation techniques,'' in \emph{Proc. Sixth Conf. Num. Sim. Plasmas},
  Berkeley, CA, 1973, pp. 46--48.

\bibitem{esirkepov2001exact}
T.~Esirkepov, ``Exact charge conservation scheme for particle-in-cell
  simulation with an arbitrary form-factor,'' \emph{Computer Physics
  Communications}, vol. 135, no.~2, pp. 144--153, 2001.

\bibitem{Vay2013PSATD}
J.-L. Vay, I.~Haber, and B.~B. Godfrey, ``A domain decomposition method for
  pseudo-spectral electromagnetic simulations of plasmas,'' \emph{Journal of
  Computational Physics}, vol. 243, pp. 260--268, 2013.

\bibitem{godfrey1974numerical}
B.~B. Godfrey, ``Numerical cherenkov instabilities in electromagnetic particle
  codes,'' \emph{Journal of Computational Physics}, vol.~15, no.~4, pp.
  504--521, 1974.

\bibitem{Godfrey2013PSATD}
B.~B. Godfrey, J.-L. Vay, and I.~Haber, ``Numerical stability analysis of the
  pseudo-spectral analytical time-domain pic algorithm,'' 2013.

\bibitem{Langdon1970}
A.~B. Langdon, ``Nonphysical modifications to oscillations, fluctuations, and
  collisions due to space-time differencing,'' in \emph{Proceedings of the
  fourth conference on numerical simulation of plasmas}, 1970, pp. 467--495.

\bibitem{Okuda1970}
H.~Okuda, ``Nonphysical instabilities in plasma simulation due to small debye
  length,'' in \emph{Proceedings of the fourth conference on numerical
  simulation of plasmas}, 1970, pp. 511--525.

\bibitem{BirdsallLangdon}
C.~K. Birdsall and A.~B. Langdon, \emph{Plasma physics via computer
  simulation}.\hskip 1em plus 0.5em minus 0.4em\relax Taylor \& Francis, 2005.

\bibitem{Mathematica9}
\BIBentryALTinterwordspacing
Mathematica, version nine. Wolfram Research Inc. [Online]. Available:
  \url{http://www.wolfram.com/mathematica/}
\BIBentrySTDinterwordspacing

\bibitem{Warp}
D.~Grote, A.~Friedman, J.-L. Vay, and I.~Haber, ``The warp code: modeling high
  intensity ion beams,'' in \emph{AIP Conference Proceedings}, no. 749, 2005
  2005, pp. 55--58.

\bibitem{godfrey2013esirkepov}
B.~B. Godfrey and J.-L. Vay, ``Numerical stability of relativistic beam
  multidimensional pic simulations employing the esirkepov algorithm,''
  \emph{Journal of Computational Physics}, vol. 248, pp. 33--46, 2013.

\bibitem{VayJCP2011}
J.-L. Vay, C.~G.~R. Geddes, E.~Cormier-Michel, and D.~P. Grote, ``Numerical
  methods for instability mitigation in the modeling of laser wakefield
  accelerators in a lorentz-boosted frame,'' \emph{Journal of Computational
  Physics}, vol. 230, no.~15, pp. 5908--5929, Jul. 2011.

\bibitem{VayPoP2011}
J.-L. Vay, C.~G.~R. Geddes, E.~Esarey, C.~B. Schroeder, W.~P. Leemans,
  E.~Cormier-Michel, and D.~P. Grote, ``Modeling of 10 gev-1 tev laser-plasma
  accelerators using lorentz boosted simulations,'' \emph{Physics of Plasmas},
  vol.~18, 2011.

\bibitem{Birdsall1980}
C.~K. Birdsall and N.~Maron, ``Plasma self-heating and saturation due to
  numerical instabilities,'' \emph{Journal of Computational Physics}, vol.~36,
  pp. 1--19, 1980.

\bibitem{VayPoPL2011}
J.-L. Vay, C.~G.~R. Geddes, E.~Cormier-Michel, and D.~P. Grote, ``Effects of
  hyperbolic rotation in minkowski space on the modeling of plasma accelerators
  in a lorentz boosted frame,'' \emph{Physics of Plasmas}, vol.~18, p. 030701,
  2011.

\bibitem{lewis1972variational}
H.~R. Lewis, ``Variational algorithms for numerical simulation of collisionless
  plasma with point particles including electromagnetic interactions,''
  \emph{Journal of Computational Physics}, vol.~10, no.~3, pp. 400--419, 1972.

\bibitem{Langdon1973energy}
A.~B. Langdon, ``Energy-conserving plasma simulation algorithms,''
  \emph{Journal of Computational Physics}, vol.~12, no.~2, pp. 247--268, 1973.

\end{thebibliography}

\end{document}